\documentclass[seceq]{ptptex}

\usepackage{graphicx}
\def\simle{%  ``less than about'' symbol
    \mathrel{\rlap{\raise 0.511ex
        \hbox{$<$}}{\lower 0.511ex \hbox{$\sim$}}}}

\title{%        %You can use \\ for explicit line-break
Latest results from lattice QCD for the Roper resonance%
}

\author{%       %Use \scshape  for the family name
Shoichi \textsc{Sasaki}%
}

\inst{%         %Affiliation, neglected when [addenda] or [errata]
Department of Physics, University of Tokyo, Tokyo 113-0033, Japan
}

\abst{%         %this abstract is neglected when [addenda] or [errata]
The present status of the Roper resonance in lattice QCD
is reviewed. Some of the latest lattice results are discussed 
with particular emphasis on a large systematic error stemming
from the finite size effect. These results suggest that the Roper resonance 
can be described by  the simple three quark excitation of sizable extent.
}

\begin{document}

\maketitle

\section{Introduction}
In the excited baryon spectrum, of particular interest is
the level order of the positive-parity excited nucleon,
so-called the Roper resonance $N'(1440)$ and 
the negative-parity nucleon $N^*(1535)$. 
It is worth mentioning that this pattern of the level order 
between positive and negative-parity excited states can be
 found universally in the $\Delta$, $\Sigma$ and flavor-octet 
$\Lambda$ channels. It is also interesting to note that 
quark confining models such as either the harmonic oscillator 
quark model or  the MIT bag model 
have some difficulty reproducing the correct level order,
as eigenstates in each model alternate in parity~\cite{Sasaki:2001nf}.
This wrong ordering problem does not seem to be
easily alleviated. Hence, the first principle calculation, 
lattice QCD is demanded to resolve such a long-standing puzzle.

\begin{figure}[th]
%\begin{minipage}[t]{155mm}
\begin{center}
\includegraphics[scale=0.4]{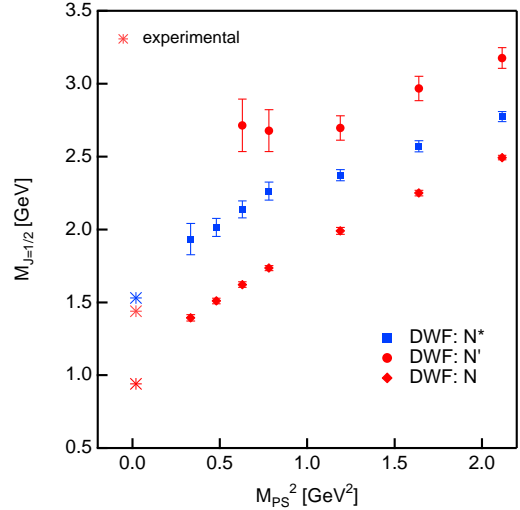}
%\vspace*{-8mm}
\caption{The low-lying nucleon spectrum in both positive parity and negative parity channel.
Note the large mass splitting of $N$ (diamonds) and $N^*$ (squares)
which is within 15\% (in the chiral limit) of the experimental value (burst). 
The mass of the positive parity excited state  $N'$ (circle) is
too high, however. All data are from Ref. 1  %\protect\cite{Sasaki:2001nf} 
where the domain
wall fermions are utilized.
The corresponding experimental value for $N(940)$, $N'(1440)$ and 
$N^*(1530)$ are marked with lower, middle and upper stars.}
\label{fig:DWF}
%\vspace*{-3mm}
\end{center}
%\end{minipage}
%\hspace{\fill}
\end{figure}

In lattice QCD, computations of the hadron masses rely on the
asymptotic behavior of the two-point hadron correlator in the large
Euclidean time region as $G(t)\sim \exp (-M_{0} t)$.
This approach mainly accesses the mass of  the lowest-lying states, $M_{0}$.
As for the higher-lying states, we need more sophisticated analysis such as
the matrix-correlator method~\cite{{Luscher:1990ck},{Allton:wc}}. 
The old calculation~\cite{Allton:wc} suggested the large
mass of the first positive-parity excited nucleon and reported the
possibility of missing the Roper resonance in lattice calculation.
Recently, a first systematic calculation for the nucleon excited states 
in both parity channels showed that the wrong ordering 
between $N'$ and $N^*$ actually happens in the relatively {\it 
heavy-quark} mass region~\cite{{Sasaki:2001nf},{Sasaki:1999yh}}, 
whereas the $N^*$ state can be reproduced well in lattice QCD~\cite{{Sasaki:2001nf},{Gockeler:2001db}} (see Fig. \ref{fig:DWF}).
What one can see in this figure closely resembles the wrong ordering problem
of the excited nucleon spectrum in the confining models.
After this work, many lattice calculations confirmed 
this particular issue.~\cite{{Lee:2000hh},{Melnitchouk:2002eg},{Maynard:2002ys}}

The question arises whether or not lattice calculations 
fail to reproduce the Roper resonance. 
An answer to this question provides some perspective to understand
the structure of the mysterious Roper resonance since above lattice calculations basically utilize the simple description of the Roper resonance as three valence quarks. However, we should not conclude anything from the previous lattice results
without examining possible systematic errors. 

From the phenomenological point of view, the mass splitting between the ground state
and the radial excited state, which has the same quantum number of the ground state, 
is almost independent of the quark masses.
For instance, such a feature is easily confirmed among spin-1/2 octet baryons as
$M_{_{N(1440)}}-M_{_{N(940)}}\approx M_{_{\Sigma(1660)}}-M_{_{\Sigma(1190)}}\approx
M_{_{\Lambda(1600)}}-M_{_{\Lambda(1115)}}
\sim 0.5 {\rm GeV}$.
In the case of the vector mesons, charm and bottom sectors
reveal more clear evidence such as:
$M_{_{\rho(1450)}}-M_{_{\rho(770)}}\approx 
M_{_{\phi(1680)}}-M_{_{\phi(1020)}}\approx M_{_{\psi(3690)}}-M_{_{J/\psi(3100)}}
\approx M_{_{\Upsilon(10020)}}-M_{_{\Upsilon(9460)}}\sim 0.6-0.7 {\rm GeV}$.

%More clearly, the following evidence can be found in  the case of vector mesons including
%charm and bottom sectors: $M_{_{\rho(1450)}}-M_{_{\rho(770)}}\approx 
%M_{_{\phi(1680)}}-M_{_{\phi(1020)}}\approx M_{_{\psi(3690)}}-M_{_{J/\psi(3100)}}
%\approx M_{_{\Upsilon(10020)}}-M_{_{\Upsilon(9460)}}\sim 0.6-0.7 {\rm GeV}$.
%
%For instance, it is the fact that $M_{_{N(1440)}}-M_{_{N(940)}}\approx %M_{_{\Sigma(1660)}}-M_{_{\Sigma(1190)}}\approx
%M_{_{\Lambda(1600)}}-M_{_{\Lambda(1115)}}
%\sim 0.5 {\rm GeV}$. 
%More clearly, one can find that $M_{_{\rho(1450)}}-M_{_{\rho(770)}}\approx 
%M_{_{\phi(1680)}}-M_{_{\phi(1020)}}\approx M_{_{\psi(3690)}}-M_{_{J/\psi(3100)}}
%\approx M_{_{\Upsilon(10020)}}-M_{_{\Upsilon(9460)}}\sim 0.6-0.7 {\rm GeV}$
%in the case of the vector meson. 
%
On the contrary, Fig. \ref{fig:DWF} shows that
the data points corresponding to the two lightest quark masses for 
$N'$ seem to behave against this empirical fact. 
In addition, the mass splitting between $N$ and $N'$ at three heaviest quark mass 
points is roughly consistent with experiment value $\approx 0.5 {\rm GeV}$.
Indeed, authors of Ref. 1 %~\cite{Sasaki:2001nf} 
remarked
that their utilized matrix-correlator method is no longer helpful in the lighter quark mass region
due to systematic uncertainties from non-negligible higher-lying contribution.

We also remark that the simulation for the light-quark mass requires
large lattice size since the ``wave function'' of 
the bound state enlarges as the quark mass decreases.
Once the ``wave function'' is squeezed due to the small volume,
the kinetic energy of internal quarks increases and thus
the total energy of the bound state should be pushed up.
This is an intuitive picture for the finite size effect on the 
mass spectrum. Such effect is expected to become serious for 
the radial excited state rather than the ground state.
Indeed, Fig. \ref{fig:DWF} shows that
the $N'$ mass in the light-quark region 
is significantly heavier than the mass 
extrapolated from the heavy quark region.
Their lattice simulation was performed on relatively 
small volume ($La\approx 1.5{\rm fm}$).

Unless taking into account these possible systematic errors,
namely the effect of high-lying states and that of finite volume, we cannot rule
out the possibility of level switching between $N'$ and $N^*$ near the chiral limit. 
To resolve the remaining puzzle, 
we utilize the maximum entropy method (MEM), instead of the 
conventional analysis such as the matrix-correlator method. 
The MEM analysis can take into account non-negligible contribution from 
higher-lying states through the reconstruction of the spectral functions (SPFs),  
$A(\omega)$ from given Monte Carlo data of the two-point hadron correlater 
$G(t)=\int d\omega A(\omega) \exp (-\omega t)$. 
Recently the MEM analysis is widely employed on various
problems in lattice simulations after the first success
in our research area \cite{{Nakahara:1999vy},{Asakawa:2000tr}}. 
Of course, we should also perform the finite size study for evaluating 
another systematic error.

\section{Latest lattice results}

Our numerical simulations are performed on three different lattice sizes, 
$L^3\times T=16^3 \times 32$, $24^3 \times 32$ and $32^3 \times 32$, 
 to determine the finite size effect and to take the infinite volume limit for
the observed masses. We generate quenched QCD configurations with the standard 
single-plaquette action at $\beta = 6.0$ ($a^{-1}\approx 1.9 {\rm GeV}$).
The quark propagators are computed using the Wilson 
fermion action at four values of the hopping parameter $\kappa$,
which cover the range $M_{\pi}\approx 0.6 - 1.0 {\rm GeV}$.
%$M_{\pi}/M_{\rho}\approx 0.69-0.92$. 
Our preliminary results are analyzed on 444 configurations 
for the smallest lattice ($La\approx1.5{\rm fm}$), 350 configurations for
the middle size lattice ($La\approx2.2{\rm fm}$) and 200 configurations 
for the largest lattice ($La\approx3.0{\rm fm}$).

%Operator
We use the conventional nucleon interpolating operators, 
$\varepsilon_{a b c}(u^{T}_{a}C\gamma_{5}d_{b})u_{c}$.
Correlators constructed from those operators are supposed to receive 
contributions from both positive and negative-parity states. More details 
of the parity projection are described in Ref. 1. %\cite{Sasaki:2001nf}. 
In this article, we focus only on the {\it positive parity} states.
Details of the MEM analysis can be found in Ref. 11. 
%~\cite{Sasaki:2002sj}.

\begin{figure}[th]
%\begin{minipage}[t]{155mm}
\begin{center}
\includegraphics[scale=0.4]{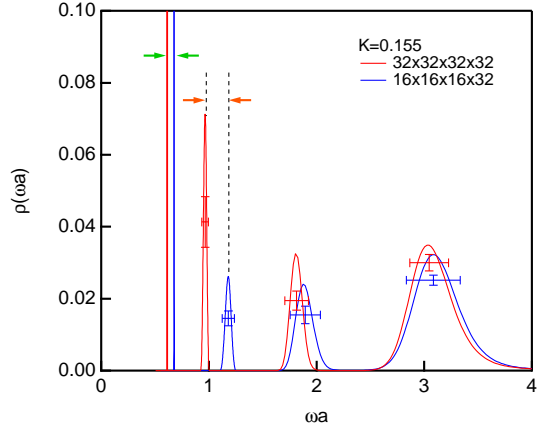}
%\vspace*{-8mm}
\caption{The dimensionless spectral function
$\rho(\omega)=A(\omega)\omega^5$
in the nucleon channel as function of the frequency $\omega$ in lattice units.
The second  peak corresponding to the first excited state is 
significantly affected by the finite size effect in comparison to the first peak 
(the ground state).}
\label{fig:SPFs}
%\vspace*{-3mm}
\end{center}
%\end{minipage}
%\hspace{\fill}
\end{figure}

First we show dimensionless spectral functions %$\rho(\omega)=A(\omega)\omega^5$ 
of  the nucleon for the smallest size (dotted line) and the 
largest size (solid line) in Fig.~\ref{fig:SPFs}. 
At glance, the gross feature in both volume is similar. 
There are two sharp peaks and two large bumps. 
The crosses on each peak or bump 
represent the statistical significance of SPF 
obtained  by the MEM.  Clearly the peak positions at each peak or bump 
are relatively shifted to the right as the lattice size decreases. The peak shift
indicates the direct observation on the finite size effect. One can see
a large finite size effect on the second peak 
as compared with the first peak and two bumps.
Here we comment that two large bumps might be the unphysical bound states 
of a physical quark and two doublers, which have been found 
in the mesonic case \cite{Yamazaki:2001er}.
We actually confirmed this speculation for baryons through the additive simulations at different lattice spacing. Those bumps appear at the same frequency $\omega$ in lattice units. 
This means that those states are infinitely heavy in the continuum limit.
We now conclude that the first two peaks are physical states but two large bumps 
are unphysical states.
Thus, Fig.~\ref{fig:SPFs} indicates that the (physical) radial excited state is 
significantly affected by the finite size effect in comparison to the ground state.

\begin{figure}[t]
\begin{center}
\begin{minipage}[t]{40mm}
\begin{center}
\includegraphics[scale=0.3]{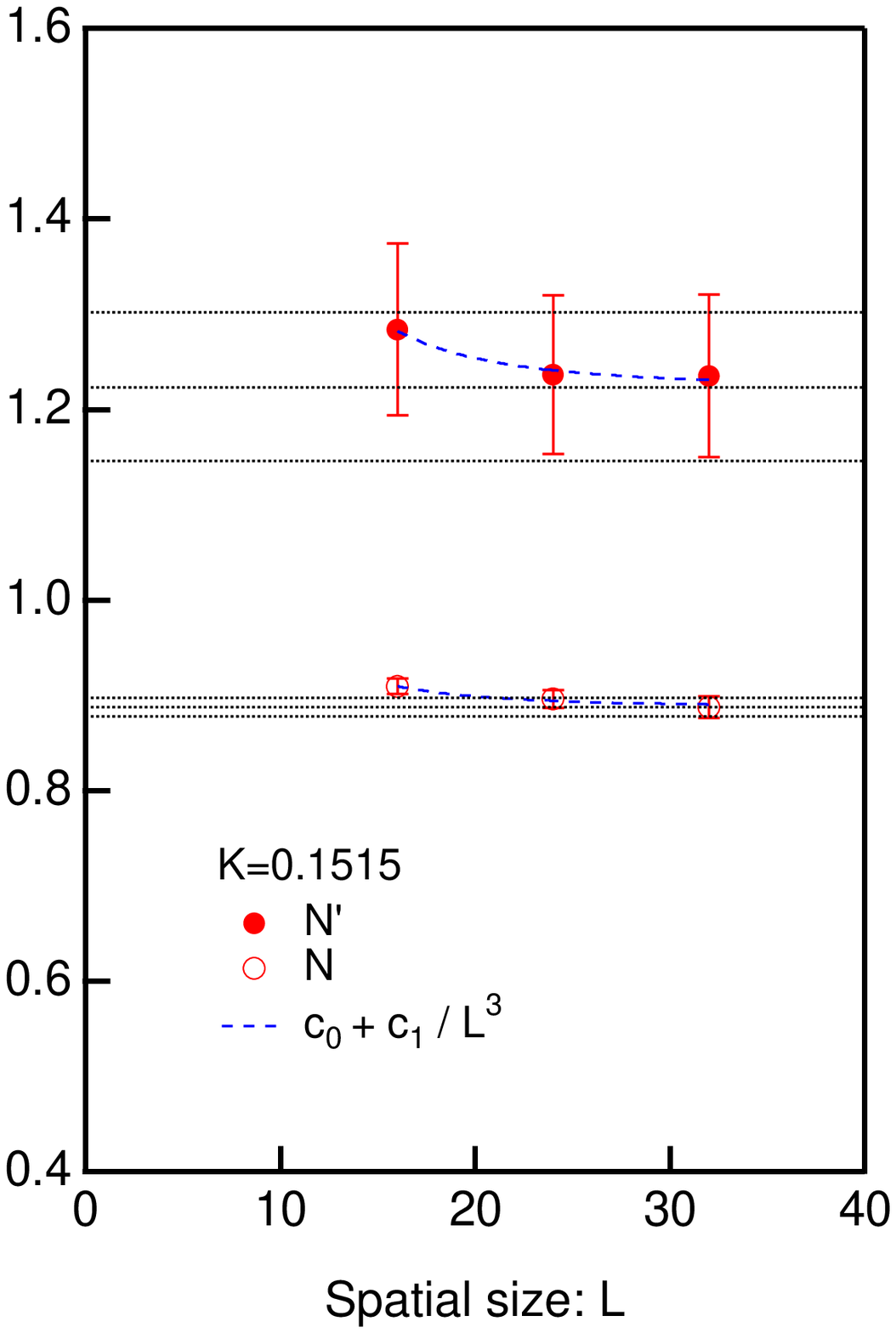}
%\vspace*{-8mm}
%\caption{caption}
\vspace*{-3mm}
\end{center}
\end{minipage}
\hspace{\fill}
\begin{minipage}[t]{40mm}
\begin{center}
\includegraphics[scale=0.3]{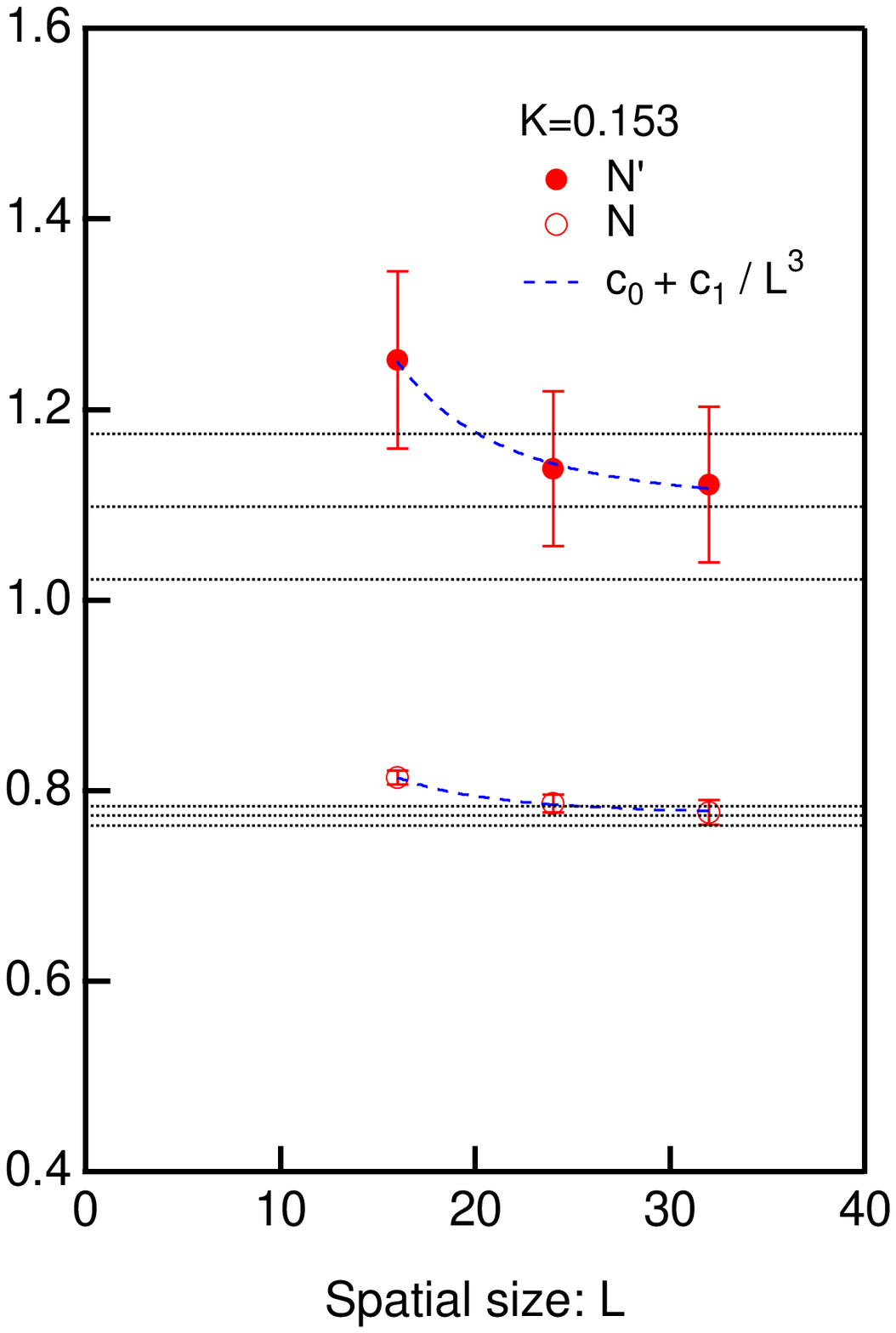}
%\vspace*{-8mm}
%\caption{caption}
\vspace*{-3mm}
\end{center}
\end{minipage}
\hspace{\fill}
\begin{minipage}[t]{40mm}
\begin{center}
\includegraphics[scale=0.3]{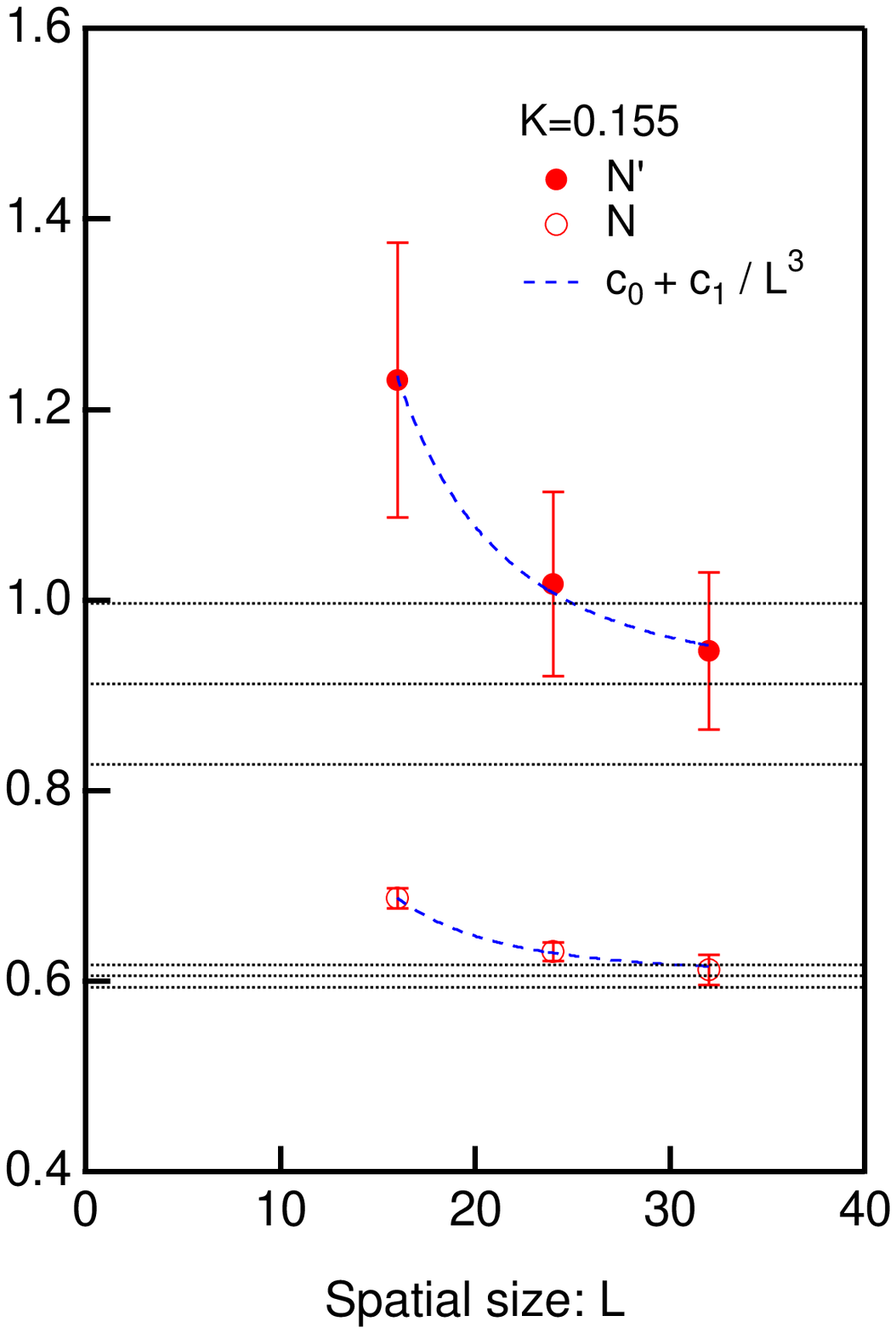}
%\vspace*{-8mm}
%\caption{caption}
\vspace*{-3mm}
\end{center}
\end{minipage}
\hspace{\fill}
\caption{Masses of the ground state (open circle)
and the first excited state (solid circle) in the 
positive-parity nucleon channel as function of the spatial lattice 
size $L$ in lattice units. Infinite volume extrapolation (dotted line) of those masses is
guided by a power law behavior.}
\label{fig:FSE}
\end{center}
\end{figure}

Next we plot the masses of the ground state (open circle) and the first excited state (solid circle), 
which correspond to the peak positions of first two peaks, as function of the spatial lattice 
size $L$ in Fig.~\ref{fig:FSE}. The errors are estimated by the jackknife method. Again, one can find the significantly large finite size effect appears in the the first excited state as compared with the ground state. Clearly, the finite-size effect is rather 
severe in the lighter quark mass region. The infinite volume limit of  the mass of each state 
at each quark mass is guided by a power law behavior as 
$M_{L}-M_{\infty}\propto c /L^3$~\cite{Fukugita:1992jj}.

\begin{figure}[t]
%\begin{minipage}[t]{155mm}
\begin{center}
\includegraphics[scale=0.4]{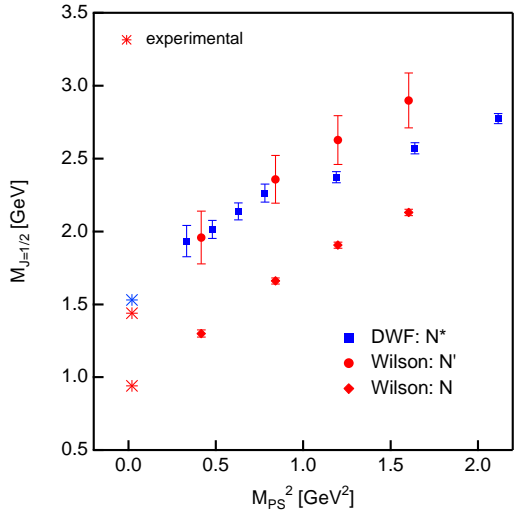}
%\vspace*{-8mm}
\caption{
Masses of the ground state (diamond) 
and the radial excited state (circle) of the nucleon in the infinite volume 
as a function of pseudo pion mass squared. 
For comparison, masses of the negative parity nucleon (square)
from Ref. 1 %\protect\cite{Sasaki:2001nf} 
are also plotted. }
\label{fig:InfwDWF}
%\vspace*{-3mm}
\end{center}
%\end{minipage}
%\hspace{\fill}
\end{figure}

Finally, we show the masses of the nucleon (diamonds) and the Roper state
(circles) in the infinite volume 
as function of the pseudo pion mass squared in Fig.~\ref{fig:InfwDWF}.
Square symbols denote the negative parity nucleon, of which data is borrowed from the 
previous DWF calculation for your eyes's guide. An important observation is that
the mass splitting between the ground state and its radial excited state does not seem
to depend on the the pseudo pion mass squared as expected. Clearly, 
the level switching between the negative parity nucleon and the corresponding Roper state 
reveal after removing the possible large systematic errors from non-negligible higher-lying contribution and the large finite size effect. In addition, the finite size effect on
the negative parity nucleon state is known to be relatively small~\cite{Gockeler:2001db}.

Recently, F.X. Lee and his collaborators %Kentucky group
also reported that a dramatic level switching between the negative parity nucleon and the
corresponding Roper state is found  in their calculation where the lightest pion mass is achieved to be close to the physical pion mass 
by using overlap fermions~\cite{Lee:2002gn}. However, there are several essential 
difference between our results and their results. They observed the strong quark mass
dependences on the mass splitting between the ground state and the radial excited state, 
which is against the empirical fact as well as our results. The level switching in their calculation takes place in the region of $M_{\pi}\sim 0.3$ to 0.4 GeV, which is far from our observation $M_{\pi}\sim 0.6$ GeV. Finally, we remark that our lightest mass for the radial excited 
state is still below the threshold of the quenched  "$\eta'$-$N$" intermediate state, which may 
contribute a dramatic spectral change such as a non-analytic behavior as function of the pion
mass squared.

\section{Summary}
We explored the level order of the positive-parity excited nucleon and the negative-parity
nucleon. Based on the systematic analysis utilizing three different lattice size, 
we confirmed the large finite size effect on the Roper state 
in the light quark mass region originally pointed out in Ref. 1. %~\cite{Sasaki:2001nf}.
The level switching between the $N^*$ state and the Roper state
should happen in lattice simulations with large spatial size which is larger than 3.0 fm.
Our lattice calculation suggests that the Roper resonance can be described by  the simple
three quark excitation of sizable extent.

\section*{Acknowledgements}
It is a pleasure to acknowledge K. Sasaki, T. Hatsuda and M. Asakawa 
for a collaboration on the subject discussed here.
%The latest results in this article are largely based on their work.
%It is a pleasure to thank my collaborators K. Sasaki, T. Hatsuda and M. Asakawa 
%Many thanks to the members of the MELQCD collaboration. The results presented in
%this article are largely based on their work.
This work is supported by the Supercomputer Project No.85 
 (FY2002) of High Energy Accelerator Research Organization (KEK).
S.S. thanks for the support by JSPS Grand-in-Aid for 
Encouragement of Young Scientists (No.13740146).	 

%\appendix
%\section{First Appendix} %Empty argument \section{} yields `Appendix'. 
%
%\section{Second Appendix}

\end{document}